\documentclass[apj]{emulateapj}

\def\ltsima{$\; \buildrel < \over \sim \;$}
\def\simlt{\lower.5ex\hbox{\ltsima}}
\def\gtsima{$\; \buildrel > \over \sim \;$}
\def\simgt{\lower.5ex\hbox{\gtsima}}
\def\kms{{\rm\,km\,s^{-1}}}

\def\kpc{{\rm\,kpc}}

\def\msun{{\rm\,M_\odot}}
\def\lsun{{\rm\,L_\odot}}

\def\pc{{\rm\,pc}}

\def\deg{^\circ}

\def\s{\ifmmode \widetilde \else \~\fi}
\def\={\overline}

\def\spose#1{\hbox to 0pt{#1\hss}}

\def\lta{\mathrel{\spose{\lower 3pt\hbox{$\mathchar"218$}}
     \raise 2.0pt\hbox{$\mathchar"13C$}}}
\def\gta{\mathrel{\spose{\lower 3pt\hbox{$\mathchar"218$}}
     \raise 2.0pt\hbox{$\mathchar"13E$}}}
\def\Dt{\spose{\raise 1.5ex\hbox{\hskip3pt$\mathchar"201$}}}    
\def\dt{\spose{\raise 1.0ex\hbox{\hskip2pt$\mathchar"201$}}}    

\def\dotsfill{\leaders\hbox to 1em{\hss.\hss}\hfill}

\def\FeH{{\rm[Fe/H]}}

\shorttitle{Deep LBT photometry of CVnI}
\shortauthors{N. F. Martin et al.}

\begin{document}


\title{A deep Large Binocular Telescope view of the Canes Venatici~I dwarf galaxy}


\author{Nicolas F. Martin$^2$, Matthew G. Coleman$^2$, Jelte T. A. De Jong$^2$, Hans-Walter Rix$^2$, Eric F. Bell$^2$, David J. Sand$^{3,4}$, John M. Hill$^5$, David Thompson$^5$, Vadim Burwitz $^6$, Emanuele Giallongo$^7$, Roberto Ragazzoni$^8$, Emiliano Diolaiti$^9$, Federico Gasparo$^{10}$, Andrea Grazian$^7$, Fernando Pedichini$^7$, Jill Bechtold$^3$}
\email{martin@mpia-hd.mpg.de}


\altaffiltext{1}{Based on data acquired using the Large Binocular Telescope (LBT). The LBT is an international collaboration among institutions in the United States, Italy and Germany. LBT Corporation partners are: The University of Arizona on behalf of the Arizona university system; Istituto Nazionale di Astrofisica, Italy; LBT Beteiligungsgesellschaft, Germany, representing the Max-Planck Society, the Astrophysical Institute Potsdam, and Heidelberg University; The Ohio State University, and The Research Corporation, on behalf of The University of Notre Dame, University of Minnesota and University of Virginia.}
\altaffiltext{2}{Max-Planck-Institut f\"ur Astronomie, K\"onigstuhl 17, D-69117 Heidelberg, Germany}
\altaffiltext{3}{Steward Observatory, The University of Arizona, Tucson, AZ 85721, USA}
\altaffiltext{4}{Chandra Fellow}
\altaffiltext{5}{Large Binocular Telescope Observatory, University of Arizona, 933 N Cherry Avenue, Tucson, AZ 85721, USA}
\altaffiltext{6}{Max-Planck-Institut f\"ur Extraterrestrische Physik, P.O. Box 1312, D-85741 Garching, Germany}
\altaffiltext{7}{INAF, Osservatorio Astronomico di Roma, via di Frascati 33, I-00040 Monteporzio, Italy}
\altaffiltext{8}{INAF, Osservatorio Astronomico di Padova, vicolo dell'Osservatorio 5, I-35122 Padova, Italy}
\altaffiltext{9}{INAF, Osservatorio Astronomico di Bologna, via Ranzani 1, I-40127 Bologna, Italy}
\altaffiltext{10}{INAF, Osservatorio Astronomico di Trieste, via G.B. Tiepolo, 11, I-34131 Trieste, Italy}

\begin{abstract}
We present the first deep color-magnitude diagram of the Canes Venatici~I (CVnI) dwarf galaxy from observations with the wide field Large Binocular Camera on the Large Binocular Telescope. Reaching down to the main-sequence turnoff of the oldest stars, it reveals a dichotomy in the stellar populations of CVnI: it harbors an old ($\gta 10$~Gyr), metal-poor ($\FeH\sim-2.0$) and spatially extended population along with a much younger ($\sim1.4-2.0$~Gyr), 0.5\,dex more metal-rich, and spatially more concentrated population. These young stars are also offset by $64_{-20}^{+40}\pc$ to the East of the galaxy center. The data suggest that this young population, which represent $\sim3-5\%$ of the stellar mass of the galaxy within its half-light radius, should be identified with the kinematically cold stellar component found by \citet{ibata06}. CVnI therefore follows the behavior of the other remote MW dwarf spheroidals which all contain intermediate age and/or young populations: a complex star formation history is possible in extremely low-mass galaxies.
\end{abstract}

\keywords{galaxies: individual (Canes Venatici I) -- galaxies: stellar content -- Local Group}

\section{Introduction}
Although the Canes Venatici~I dwarf galaxy (CVnI; $\alpha_0=13^\mathrm{h}28^\mathrm{m}03.5^\mathrm{s}$, $\delta_0=33\deg33'21.0''$) is not much less luminous than previously known satellites of the Milky Way ($M_V=-7.9\pm0.5$), its large distance ($224\pm21\kpc$) kept it hidden until it was recently discovered by \citet{zucker06a} in the Sloan Digital Sky Survey (SDSS). Unlike most other Galactic satellites at more than $200\kpc$ from the MW, CVnI seems dominated by old stellar populations. However, applying an automated color-magnitude diagram (CMD) fitting technique to SDSS data, \citet{dejong07} show tentative evidence for the presence of a much younger population of only $\sim3$~Gyr.

A spectroscopic survey of CVnI stars \citep{ibata06,martin07a} also revealed the presence of two kinematically distinct populations. The most metal-rich half of the sample ($-1.9\lta\FeH\lta-1.5$) shows a kinematically extremely cold component, with a radial velocity dispersion $<1.9\kms$, whereas the most metal-poor half of the sample ($-2.5\lta\FeH\lta-1.9$) is much hotter ($\sigma\gta10\kms$). The hot component is also measured to be twice as spatially extended as the cold one. However these findings have been challenged by \citet{simon07} who do not find such a dichotomy of kinematic properties in their larger spectroscopic sample. 

In order to better understand the stellar populations that are present in CVnI and understand their role in shaping the structure of the dwarf galaxy, we have used the blue Large Binocular Camera (LBC, \citealt{ragazzoni06}, Giallongo et al. in prep.) on the Large Binocular Telescope (LBT, \citealt{hill06}) to obtain photometric observations that are deep enough to reach the main sequence turnoff (MSTO) of the oldest stars. We briefly present the data in section~2 before analyzing and discussing them in section~3 and~4.

\section{Data}

The observations comprise $6\times5$\,min and $4\times5$\,min exposures in the B and V band respectively and were performed as part of the LBC-blue Science Demonstration Time program during the night of June 11, 2007 with a seeing of $1.1-1.4$\,arcsec. Data reduction was handled with the same pipeline as used for the reduction of the Hercules dwarf galaxy LBT observations \citep{coleman07}.

A comparison of the LBT photometry ($B_\mathrm{LBT}$ and $V_\mathrm{LBT}$) with the SDSS stars having $g<22.5$, transformed into Landolt $B$ and $V$ \citep{jester05}, reveals the presence of a slight color term that we correct by its best linear fit:

\begin{displaymath}
V =  V_\mathrm{LBT} + 0.051 \cdot (B-V)_\mathrm{LBT} - 0.121.
\end{displaymath}

\noindent An iterative $3\sigma$ clipped Gaussian fit of the residuals around this linear correction yields a dispersion of 0.08\,mag. $B-B_\mathrm{LBT}$ is found to be close to zero (although with a larger scatter) and no correction is applied. While this is not optimal, it will not impact our conclusions that are dependent on imaging depth rather than photometric precision.

The 50\% completeness limits are $B_{50}=25.90\pm0.07$ and $V_{50}=26.00\pm0.08$ and the $5\sigma$ flux limits are reached for $B=25.2$ and $V=25.1$. Finally, we only keep stars with $\delta m<0.15$ in both magnitudes and with a DAOPHOT sharpness parameter in the B band of $\pm0.25$.

\section{Analysis}
\subsection{Revised distance from the horizontal branch}

The color-magnitude diagram of the region within the half-light radius ($r_{hb}=8.4'$, ellipticity $= 0.38$ and PA $= 73\deg$; \citealt{zucker06a}) of CVnI is shown in the left panel of Fig.~1 and provides a quite dramatic improvement over the shallow SDSS CMD that barely reaches the horizontal branch at $V\sim22.5$ (HB, \citealt{zucker06a}). Panel \emph{b} shows the underlying CMD contamination in a region of equal coverage at the Southern edge of the field. Given the good LBT photometry (the errors at these magnitudes are dominated by the dispersion in the color correction from \S~2), these stars can be used directly to constrain the distance of the dwarf galaxy more precisely than from the tip of its sparsely populated red giant branch (RGB). The median magnitude of these HB stars is  $m_V(\textrm{HB}) = 22.22 \pm 0.09$. To determine the absolute magnitude of these stars, we follow \citet{cacciari03}:

\begin{displaymath}
M_V(\textrm{HB}) = (0.23\pm0.04) \cdot (\FeH+1.5) + (0.59\pm0.03)
\end{displaymath}

As the spectroscopic sample from \citet{ibata06} and \citet{martin07a} also covers the region within the half-light radius, we use their median spectroscopic $\FeH=-2.0\pm0.2$ to determine $M_V(\textrm{HB}) = 0.48 \pm 0.06$. Assuming $A_V=0.05$ \citep{schlegel98,zucker06a} finally leads to $(m-M)_0=21.69\pm0.10$ or $D=218\pm10\kpc$. A value that is in agreement with, but more accurate than, previous determinations from \citet{zucker06a} and \citet{dejong07}.

\subsection{Stellar populations}

The LBT data are deep enough to show the subgiant branch of the galaxy and reach the MSTO of old populations. To remove the contamination from background compact galaxies masquerading as stars, we subtract the scaled Hess diagram of regions to the North and South of the galaxy center, where none of the CMD features of CVnI are found ($|\delta-\delta_0|>9'$, see Fig.~2). The resulting diagram is shown (identically) in panels \emph{c} to \emph{e} of Fig.~1 and reveals the turnoff of an old stellar population at $V\sim25.0$ and $0.0\lta B-V \lta 0.5$. Both the CMD and Hess diagram also reveal a blue plume (BP) of stars at $B-V\sim0.1$ and $23.5 \lta V\lta 25.0$, betraying the presence of a possible young stellar population.

The isochrones of \citet{girardi02} with an age of 14.1~Gyr (panel \emph{b}) are in good agreement with the color and magnitude of the `old red population' of CVnI, as they overlap nicely with the turnoff, the sub-giant branch and give the proper location of the horizontal branch for metallicities in the range $-2.3\lta\FeH\lta-1.7$. The relatively red color of the RGB, however, seems to indicate a metallicity above $-2.0$. Isochrones of 10.0~Gyr also give a reasonable agreement, although in this case, the MSTO barely copes with the data, suggesting that the old population of CVnI has an age $\gta10$~Gyr. 

The origin of the BP is harder to ascertain given the well known confusion between young stars and blue stragglers produced by binaries in dwarf galaxies. However, determining the blue straggler frequency $F_{\mathrm{HB}}^{\mathrm{BS}}$ as defined in \citet{momany07} yields $\log(F_{\mathrm{HB}}^{\mathrm{BS}})=0.5\pm0.2$. A value that is $\sim1.5\sigma$ away from the relation these authors find in galaxies believed to contain no young stars. Moreover, the spatial offset of the BP stars from the center of CVnI (see below) would require the binary population at the origin of the putative blue stragglers to have different properties form the bulk of the CVnI stars. Finally, the presence of a carbon-star \citep{zucker06a} as well as Cepheid variables \citep{kuehn07} in CVnI would be a natural outcome of the presence of a young/intermediate age population. All these points favor the presence of a young population in CVnI, with an age between $\sim1.4$ and $\sim2.0$~Gyr from the isochrone fitting of Fig.~1 (panel \emph{d}). These isochrones also need to be more metal-rich than the old population ($\FeH\sim-1.3$, although a slightly more metal-rich population is still in agreement with the data). In addition, they can explain the fuzz of stars in the Hess diagram at $V\sim23.0$ and $B-V\sim0.3$.

To determine the relative luminosity contribution of both components in CVnI, we extrapolate the completeness corrected luminosity measured in the two selection boxes of Fig.~1\emph{e}. Using the 2~Gyr, $\FeH=-1.5$ luminosity function from \citet{dotter07}, with a Salpeter initial mass function \citep{salpeter55}, we extrapolate a total luminosity of $L_{\mathrm{V,yng}}\sim3.5\cdot10^3\lsun$ within $r_{hb}$. The 12~Gyr, $\FeH=-2.0$ luminosity function is assumed for the old population and yields total luminosity $L_{\mathrm{V,old}}\sim3.5\cdot10^4\lsun$, once again within the half-light radius. The young stars therefore represent $\sim10\%$ of the luminosity of the galaxy, which translates to $\sim3-5\%$ of the stellar mass of the galaxy within $r_{hb}$ (although there are sizable uncertainties on this number given the assumptions made, especially on the IMF). 

\subsection{Spatial distribution of stellar populations}
To build the maps of these two populations, we use the ``matched filter'' method that is the optimal search strategy (in the least-square sense) to recover a signal, assuming that one has a perfect model of the signal and of its contamination (see e.g. \citealt{rockosi02} for more details). In this case, the signal is the CMD features of CVnI and the contamination is produced by foreground stars and background galaxies in the CMD. Contrary to a simple density map that gives the same weight to each star, the matched-filter technique assigns a higher weight to stars that belong to a high-contrast CVnI CMD feature (e.g. HB or RGB stars). This weight is defined as $N_\mathrm{sig}/N_{\mathrm{back}}$ where $N_\mathrm{sig}$ is the number of stars in the signal CMD  that fall within a $0.2\textrm{mag}\times0.2\textrm{mag}$ box centered around the considered star (and similarly for $N_{\mathrm{back}}$ and the contamination CMD). The signal CMD is the one of Fig. 1\emph{a} after removing obvious non-members with $B-V>1.5$ and the contamination CMD is obtained from regions of the LBT field with $|\delta-\delta_0|>9'$ (hatched region in top-left panel of Fig.~2).

The resulting contour maps of detections higher than the mean background level are shown in Fig.~2 for all the stars in the sample (top-left panel), and  the red and blue selection boxes of Fig.~1\emph{e} (top-right and bottom-left respectively)\footnote{For the sparser blue population that is more strongly affected by background galaxies we also apply a stricter cut of $\pm0.15$ on the DAOPHOT sharpness parameter.}. The various maps are overlaid in the bottom-right panel for a direct comparison. The galaxy exhibits rather distorted contours, although the degrading star/galaxy separation at fainter magnitudes can also be (at least partly) responsible for this effect. Shallower regions produced by the gaps between the CCDs of the camera at $(\alpha-\alpha_0)\cos(\delta_0)=\pm4'$ also seem responsible for the boxy shape of the inner regions of the dwarf. A more surprising feature is the double core at the center of CVnI, along with the slight offset of its densest region that lies to the East. Given that the old stars are distributed over the whole galaxy and are centered on ($\alpha_0$,$\delta_0$), this double-core has to be produced by the much more spatially confined young population offset by $1.0_{-0.3}^{+0.6}$\,arcmin or $64_{-20}^{+40}\pc$ to the East (the uncertainties have been determined by a bootstrap resampling of the position of the BP stars within $r_{hb}$).

The spectroscopic survey of \citet{ibata06} and \citet{martin07a} has also revealed the presence of two distinct components in this galaxy from the spectroscopy alone. The methodologically independent discovery of two distinct stellar components in CVnI from the LBC photometry and the much smaller extent of the young population surprisingly recalls their findings.

A direct comparison of their spectroscopic sample with the LBC contour maps of the old and young populations is presented in Fig.~3. Stars belonging to the kinematically cold, more metal-rich population are represented by filled (hollow) stars when they are within $1\sigma$ ($2\sigma$) of the \citet{ibata06} radial velocity peak of this component (see the top-middle panel of their Fig.~2). Other stars in the sample belong to the kinematically hot component and are plotted as red dots. These stars extend over the contours of the old population (left panel) and do not exhibit any clustering whereas the cold component stars are mainly clustered over or close to the contours of the young population (right panel). Although the numbers considered here are small and a random clustering of these stars  does not have a negligible probability, the fact that they closely follow the distribution of the \textit{independently} selected young stars suggests that the cold component of \citet{ibata06} is indeed real and that it corresponds to the young population seen in the LBT data. The hot component would then be produced by the old population of the galaxy. The metallicities between the spectroscopic and photometric data are also in agreement since the hot component is metal-poor with $\FeH\sim-2.0$ (compared to $-2.3\lta\FeH\lta-1.7$ from the LBT photometry) and the cold one more metal-rich by $\sim0.5$\,dex in both cases.

The only significant discrepancy between the two surveys is the relative proportion of the two components since the young stars should represent less than 10\% of the CVnI stars within its half-light radius. However, assuming the kinematic properties of the two populations from \citet{ibata06}, only $\sim9$ hollow or filled stars of Fig.~3 are expected to be genuine members of the cold components with the other being contaminants from the hot component. This translates to $\sim14\%$ of the spectroscopic sample. Given the small number of stars we are dealing with, the difference in the proportion of young stars in the photometry and cold stars in the spectroscopic sample is reasonable.

\section{Conclusion}
From deep B- and V-band photometry of the Canes Venatici~I galaxy reaching to the old main sequence turnoff, we found evidence for:
\begin{itemize}
\item A (known) dominant old population that represents $\sim95$ percent of the mass of the galaxy, with an age $\gta10$~Gyr and $\FeH\sim-2.0$;
\item A new, much younger population revealed by a blue plume that corresponds to $\sim1.4-2.0$~Gyr stars that are also slightly more metal-rich ($\FeH\sim-1.5$).
\end{itemize}

\noindent We show that the old population is extended and mainly shapes the structure of CVnI. In contrast the young population is very compact as well as offset from the dwarf center by $1.0_{-0.3}^{+0.6}$\,arcmin or $64_{-20}^{+40}\pc$. This suggests that it is not yet dynamically relaxed although this is not surprising given that the typical relaxation time for a dwarf galaxy is higher than a Hubble time (e.g. \citealt{gilmore07}). Such a clump could therefore be used to constrain the shape of the CVnI potential, as has been done in UMi \citep{kleyna03}.

These two distinct populations are in good agreement with previous more tentative determinations from an automated study of the SDSS CMD of CVnI by \citet{dejong07}. They are also in line with the spectroscopic results from \citet{ibata06} who have shown that the dwarf harbors two kinematically very different populations with different metallicities. From their spatial distribution, we link their extremely cold component to the young compact population that we found in the LBT data.

Even though it is a faint and low-mass galaxy ($10^7-10^8\msun$, \citealt{martin07a}, \citealt{simon07}) CVnI, that is located at a distance of $218\pm10\kpc$, behaves as its brighter counter-parts located at $\gta200\kpc$. Indeed, it exhibits un-mixed, distinct components and a complex star formation history.

\acknowledgments
The authors gratefully thank the LBT Science Demonstration Time team for making these observations possible.

{\it Facilities:} \facility{LBT (LBC-Blue)}.


\clearpage

\begin{figure*}
\begin{center}
\includegraphics[angle=270,width=\hsize]{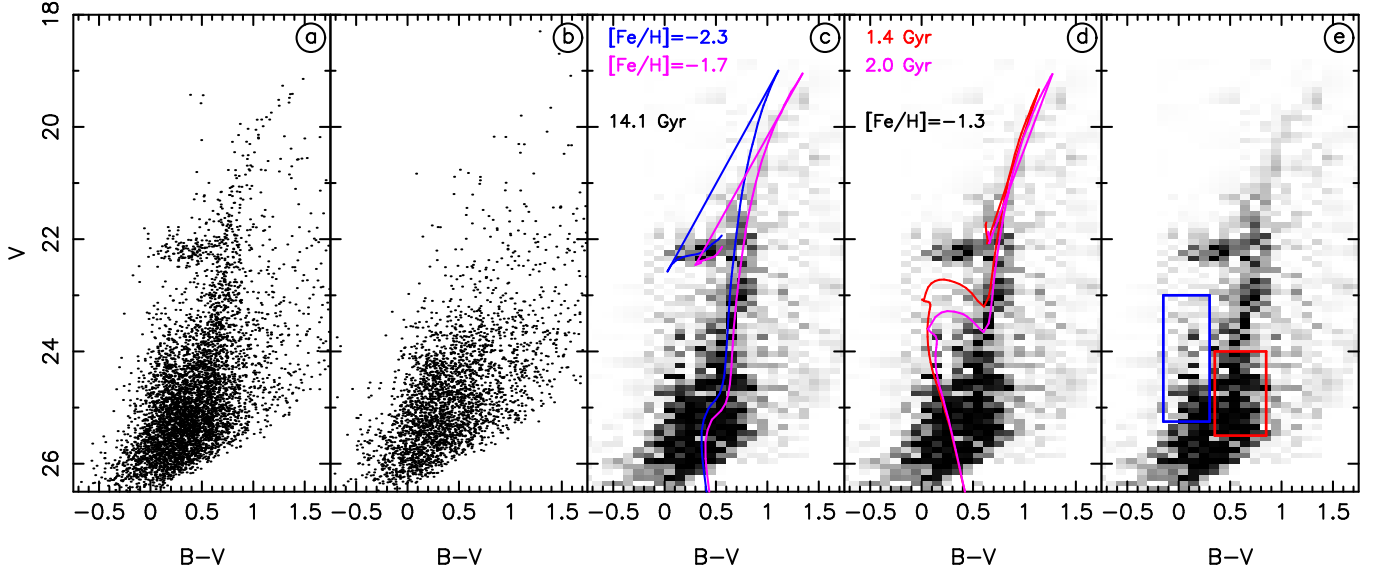}
\caption{LBC CMD within the half-light radius of the Canes Venatici~I dwarf galaxy (panel \emph{a}) and for the background population  (panel \emph{b}) over an equal area. The RGB and HB of the dwarf galaxy are clearly visible at $V<22.5$. Although fainter regions start suffering from the contamination of background compact galaxies, the MSTO of the old stars is visible at $V\sim25.0$ and $0.0\lta B-V \lta 0.5$ and a BP is also present around $B-V\sim0.1$ and $23.0 \lta V\lta 25.0$. These features are more apparent in the background-corrected Hess diagram (panels \emph{c} to \emph{e}). The \citet{girardi02} isochrones have been overlaid for metallicities of $\FeH=-2.3$ and $-1.7$ and an age of 14.1~Gyr in panel \emph{c} and match the red MSTO. On panel \emph{d}, much younger isochrones of only 1.4 and 2.0~Gyr, with $\FeH\sim-1.3$ provide a good fit to the blue population although slightly more metal-rich populations also agree with the data. These young isochrones then overlap with the galaxy's (sub)giant branch. Finally, the selection boxes used for Fig.~2 are shown on panel \emph{e}.}
\end{center}
\end{figure*}

\begin{figure}
\begin{center}
\includegraphics[angle=270,width=\hsize]{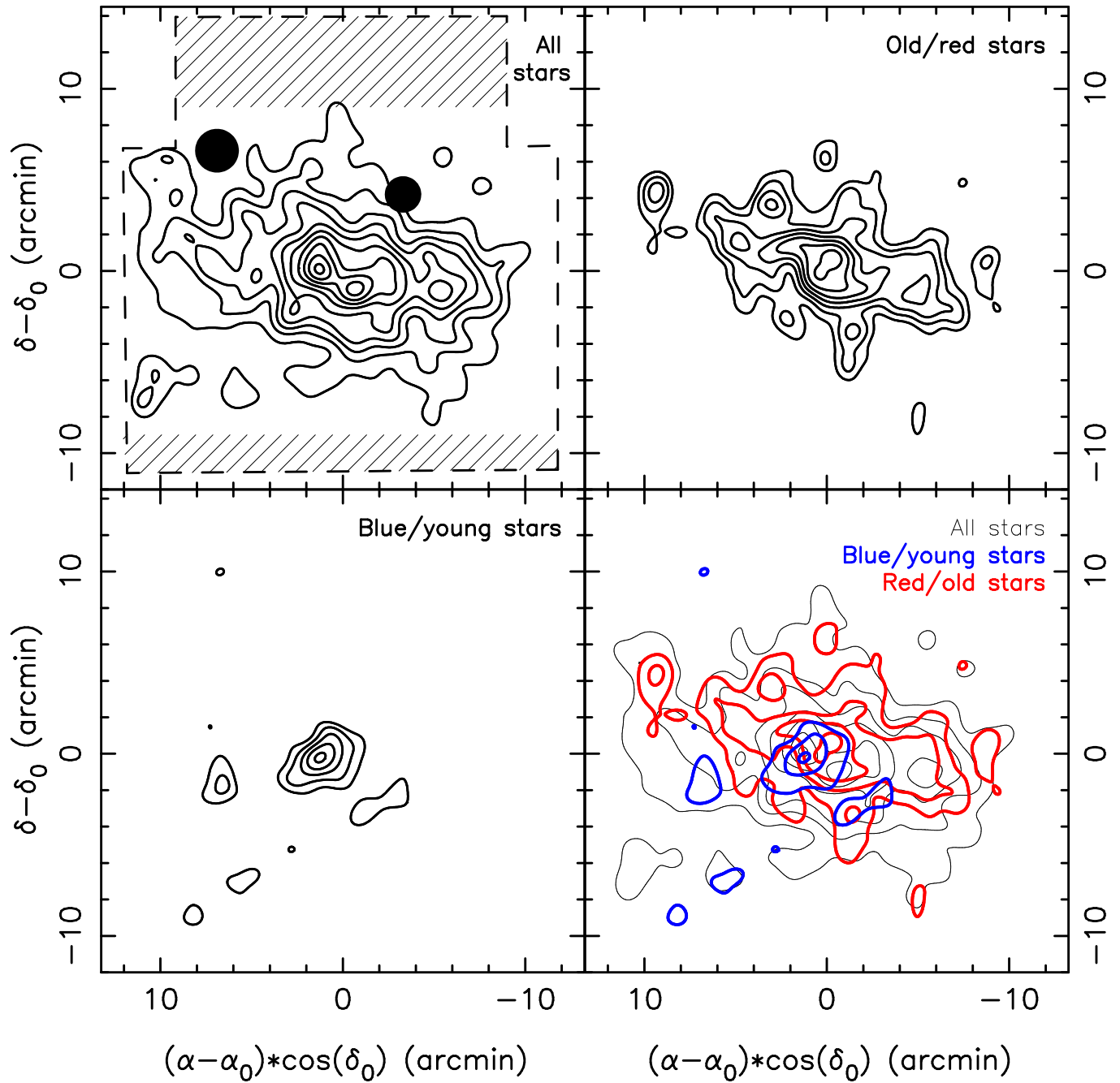}
\caption{Matched filter contour-maps of all stars in the LBT field (top-left panel). The contours correspond to regions that are $3\sigma$, $5\sigma$, $7\sigma$,\dots over the mean background density. The dashed line represents the extent of the LBT field and the hashed regions were used to determine the background contamination. The two black dots correspond to the location of two saturated foreground bright stars. The top-right and bottom-left panels show similar maps for the red/old and blue/young stars respectively, selected within the boxes of Fig.~1\emph{e}. Contours correspond here to $3\sigma$, $4\sigma$, $5\sigma$, \dots detections over the background. The young population is offset from the center of the dwarf by $1.0_{-0.3}^{+0.6}$\,arcmin and much more spatially confined than the old one. All the contour-maps are shown in the bottom-right panel for a direct comparison with all stars contours shown in black, red/old stars contours in red and blue/young stars contours in blue. To avoid cluttering the plot, only half of the contours are drawn here.}
\end{center}
\end{figure}

\begin{figure}
\begin{center}
\includegraphics[angle=270,width=0.8\hsize]{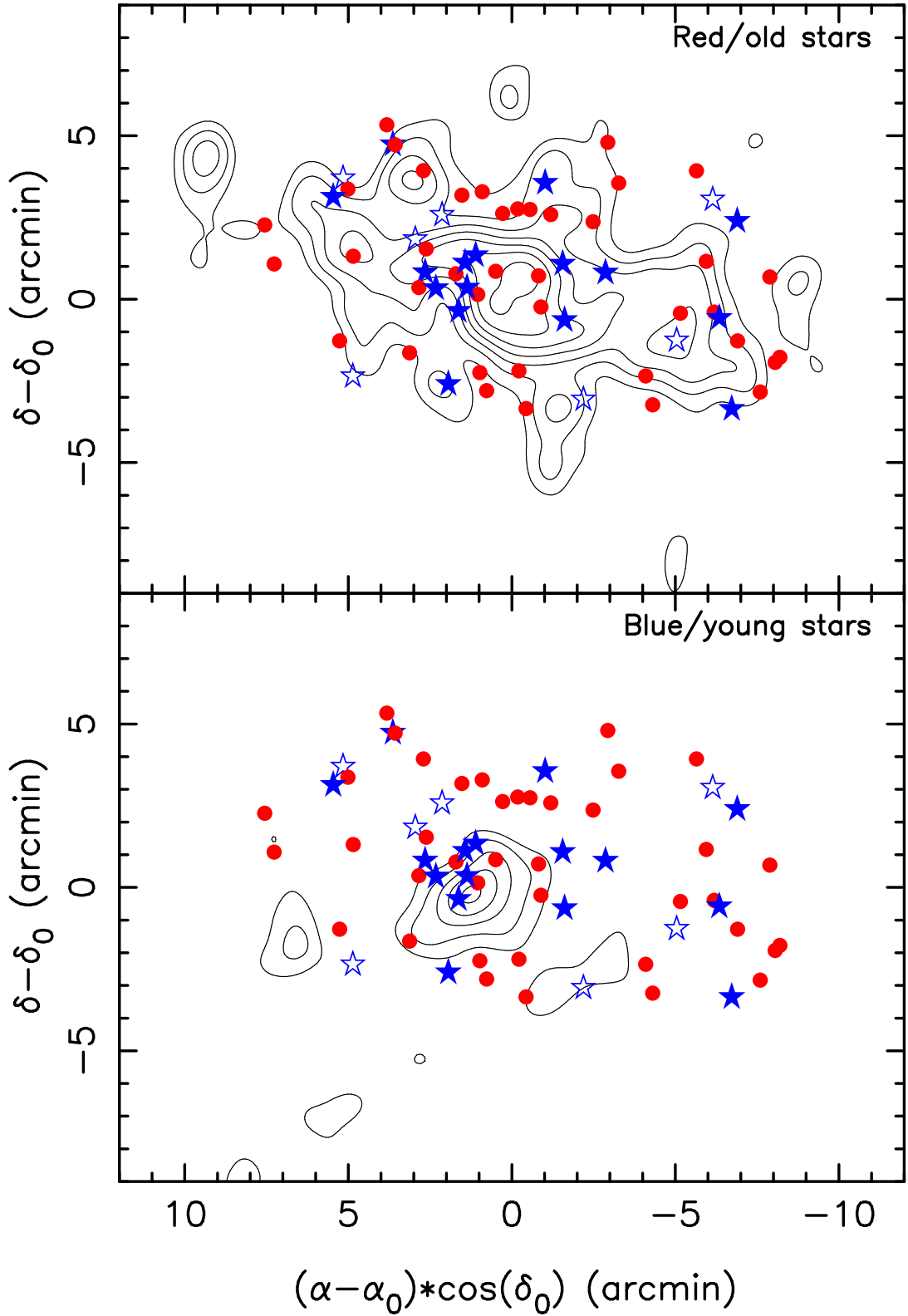}
\caption{Comparison between the LBT contour-maps for the old and young populations of CVnI (top and bottom panels respectively) and the spectroscopic sample of \citet{ibata06} and \citet{martin07a}. Stars within $1\sigma$ of the cold kinematic component are shown as filled blue stars (those within $2\sigma$ as hollow blue stars) and tend to be located within or close to the young population contours. Other stars, that belong to the hot component of CVnI, appear as red dots and show a smooth distribution over the contours of the old population.}
\end{center}
\end{figure}

\end{document}